\documentclass[twocolumn,floatfix,amsmath,amssymb,nofootinbib,superscriptaddress,prl]
{revtex4}
\pdfoutput=1

\usepackage{graphicx}
\usepackage{amsmath,amsfonts,amssymb}
\usepackage{color}
\usepackage{float}
\usepackage{epsfig,graphics,subfig,float}
\usepackage{amssymb,amsmath}
\usepackage{subfig}
\usepackage{hyperref}
\usepackage{listings}
\graphicspath{{figures/}}

\definecolor{dkgreen}{rgb}{0,0.6,0}
\definecolor{gray}{rgb}{0.5,0.5,0.5}
\definecolor{mauve}{rgb}{0.58,0,0.82}

\usepackage{tensor}
\usepackage{cancel}
\usepackage{color}
\newcommand{\HCd}{\mathcal{H}}

\def\HCdt0{\tilde{\HCd}_{0}}
\newcommand{\afffias}{Frankfurt Institute for Advanced Studies (FIAS), Ruth-Moufang-Strasse~1, 60438 Frankfurt am Main, Germany}

\newcommand{\affbgu}{Physics Department, Ben-Gurion University of the Negev, Beer-Sheva 84105, Israel}
\newcommand{\affbahamas}{Bahamas Advanced Study Institute and Conferences, 4A Ocean Heights, Hill View Circle, Stella Maris, Long Island, The Bahamas}
\newcommand{\affpoland}{Institute of Theoretical Physics, University of Wroclaw, 50-204 Wroclaw, Plac Maxa Borna 9, Poland}
\begin{document}

\title{Unification of DE - DM from Diffusive Cosmology}
\author{D. Benisty}
\email{benidav@post.bgu.ac.il}
\affiliation{\afffias}\affiliation{\affbgu}
\author{E.I. Guendelman}
\email{guendel@bgu.ac.il}
\affiliation{\afffias}\affiliation{\affbgu}\affiliation{\affbahamas}
\author{Z. Haba}
\email{zhab@ift.uni.wroc.pl}
\affiliation{\affpoland}
\date{\today}
\begin{abstract}
Generalized ideas of unified dark matter and dark energy in the context of dynamical space time theories with a diffusive transfer of energy are studied. The dynamical space-time theories are introduced a vector field whose equation of motion guarantees a conservation of a certain Energy Momentum tensor, which may be related, but in general is not the same as the gravitational Energy Momentum tensor. This particular energy momentum tensor is build from a general combination of scalar fields derivatives as the kinetic terms, and possibly potentials for the scalar field. By demanding that the dynamical space vector field be the gradient of a scalar the dynamical space time theory becomes a theory for diffusive interacting dark energy and dark matter. These generalizations produce non-conserved energy momentum tensors instead of conserved energy momentum tensors which leads at the end to a formulation for interacting DE-DM. We solved analytically the theories and we show that the $\Lambda$CDM is a fixed point of these theories at large times. A particular case has asymptotic correspondence to previously studied non-Lagrangian formulations of diffusive exchange between dark energy dark matter.    
\end{abstract}
\received{\today}
\maketitle
\section{Introduction}
Dark energy and Dark matter constitute most of the observable Universe. Yet the true nature of these two phenomena is still a mystery. One fundamental question with respect to those phenomena is the coincidence problem which is trying to explain the relation between dark energy and dark matter densities. In order to solve this problem, one approach claims that the dark energy is a dynamical entity and hope to exploit solutions of scaling or tracking type to remove dependence on initial conditions. Others left this principle and tried to model the dark energy as a phenomenological fluid which exhibits a particular relation with the scale factor \cite{Cardone:2004sq}-, Hubble constant \cite{Dvali:2003rk} or even even the cosmic time itself \cite{Basilakos:2009ah}.

Interaction between DM and DE was considered in many cases, such as \cite{Arevalo:2016epc}. Unifications between dark energy and dark matter from an action principle were obtained from scalar fields \cite{Scherrer:2004au}-\cite{Arbey:2006it} or by other models \cite{Chen:2008ft}-\cite{Leon:2013qh} including Galileon cosmology \cite{Leon:2012mt} or Telleparallel modified theories of gravity \cite{Kofinas:2014aka}-\cite{Skugoreva:2014ena}. Beyond those approaches, a unification of Dark Energy and Dark Matter using a new measure of integration (the so-called Two Measure Theories) has been formulated \cite{Guendelman:2015jii}-\cite{Guendelman:2016kwj}. A diffusive interaction between dark energy and dark matter was introduced in \cite{Koutsoumbas:2017fxp}-\cite{Haba:2016swv} and was formulated in the context of an action principle based on a generalization of those Two Measures Theories in the context of quintessential scalar fields \cite{Benisty:2017eqh}-\cite{Benisty:2017rbw}. 

In recent publications \cite{Calogero:2013zba}, diffusion of energy between dark energy into dark matter was discussed. The models of such type are interesting as an approach to solve the coincidence problem. The basis of those  models are considering a non-conserved stress energy tensor $T^{\mu\nu}$ with a source current $j^\mu$:
\begin{equation} \label{diffusion}
\nabla_\mu T^{\mu\nu}_{(\textbf{Dust})}=\gamma^{2} j^\nu
\end{equation}
where $\gamma^{2}$ is the coupling diffusion coefficient of the fluid. The current $j^\mu$ is a time-like covariant conserved vector field $j^{\mu}_{;\mu}=0$ which describes the conservation of the number of particles in the system. Due to the fact that the Einstein tensor is  covariantly conserved $\nabla_\mu G^{\mu\nu}=0$, we have to introduce on the right hand side of the Einstein tensor a compensating energy momentum tensor, for two diffusive fluids, where :
\begin{equation}\label{dedm}
	\nabla_\mu T^{\mu\nu}_{(Dust)} = -\nabla_\mu T^{\mu\nu}_{(\Lambda)} =\gamma^{2}  j^\nu
\end{equation}
so that the total energy momentum tensor is conserved:
\begin{equation}
	\nabla_\mu \left( T^{\mu\nu}_{(Dust)} +  T^{\mu\nu}_{(\Lambda)} \right)= 0
\end{equation}
Such models could originate from irreversible diffusive exchange of energy, or have a Lagrangian origin, by introducing an independent stress energy momentum tensor $T^{\mu\nu}_{(\chi)}$ directly in the Lagrangian. 
\begin{table*}[t!]
  \begin{center}
     \begin{tabular}{l|c|c|r}  
    \textbf{Name} & \textbf{The point} & \textbf{Eigenvalues} & \textbf{Densities fraction} \\
       \hline
      A & $\left(0, \frac{3}{\gamma}  (\omega -\tilde{\omega})\right)$ & $3 (3 \omega +1)$ , $ 3 (\omega -\tilde{\omega})$ & $0$ \\
      B &  $\left(\frac{\omega +1/3}{\omega -\tilde{\omega}},-\frac{3 \tilde{\omega}+1}{\gamma }\right)$  & $\frac{1}{2} \left(\pm \sqrt{36 \omega ^2-72 \omega  \tilde{\omega}+9 (\tilde{\omega}-2) \tilde{\omega}-3}-6 \omega -3 \tilde{\omega}-3\right)$ & $-\frac{3 \omega +1}{3 \tilde{\omega}+1}$ \\
      C & $\left(0,0 \right)$  & $3 ( \tilde{\omega}-\omega )$, $3 (2 \omega +\tilde{\omega}+1)$ & $0$ \\
      D & $\left(\frac{2 \omega + \tilde{\omega} + 1}{2 (\omega -\tilde{\omega})}, 0 \right)$ & $-3 (2 \omega + \tilde{\omega} +1)$, $\frac{3}{2} (1 + 3  \tilde{\omega})$ & $ -\frac{2 \omega + \tilde{\omega}+1}{3  \tilde{\omega}+1} $\\
    \end{tabular}
    \caption{The properties of the critical points for the exponential potential}
  \end{center}
   \label{tab:table1}
   \end{table*}
The structure of the paper is as follows: In section (2) we discuss dynamics of exchange of energies between two diffusive fluids, with two different equation of states. Such a system has a universal model independent behavior. In section (3) we present the Lagrangian model leading to such an interactive energy momentum tensor. In section (4) we discuss solutions for the theory which contains more general combinations for the stress energy momentum tensor $T^{\mu\nu}_{(\chi)}$. In Section (5) we are looking for few asymptotic solutions for the theory. In Section (6) we discuss a special case of a Lagrangian which corresponds to the diffusive model which has been introduced in section (2).  
\section{coupled diffusive fluids}
We assume that stress energy momentum tensors are in the form of ideal fluids, where:
\begin{equation}
T^{\mu}_{\nu} = \textbf{Diag} (\rho,-p,-p,-p)
\end{equation}
where $\rho$ is the energy density and p is the pressure. Then equations (\ref{diffusion})-(\ref{dedm}) read:
\begin{equation}
\dot\rho_{dust} + 3 H (1+\tilde{\omega}) \rho_{dust} = \frac{\gamma^2}{a^3}
\end{equation}
and
\begin{equation}
\dot\rho_{\Lambda} + 3 H (1+\omega) \rho_{\Lambda} = -\frac{\gamma^2}{a^3}
\end{equation}
The diffusion constant $\gamma^2$ is always positive. $\omega$ and $\tilde{\omega}$ denote the ratio of the pressure and the density for the corresponding fluids. In order investigate the behavior of the solution, we introduce the dynamical system method for the equations. The dimensionless quantities for the system are defined as \cite{Haba:2016swv}: 
\begin{equation}
x = \frac{\rho_{dust}}{3 H^2}, \quad y = \frac{\rho_{\Lambda}}{3 H^2}, \quad \delta = \frac{\gamma^2}{a^3 H \rho_{dust}}
\end{equation}
where $\delta$ describes the strength of the relative diffusion. From Friedmann equations $x+y = 1$.
The complete autonomous system method equations are: 
\begin{subequations}
\begin{equation}
x' = 6 x^2 (\tilde{\omega}-\omega )+ x (\gamma \delta +3 +6 \omega +3 \tilde{\omega})
\end{equation}
\begin{equation}
\delta' = \delta  (\gamma \delta +3 (x-1) (\omega -\tilde{\omega}))
\end{equation}
\end{subequations}
Table (1) presents the critical points in the system. In order to determine the stability of the system we have to specify the equations of states. For the case of dark matter and dark energy we can choose two cases: the first on: $\omega = -1, \tilde{\omega} = 0$ and the second one $\omega = 0, \tilde{\omega} = -1$. The case $\omega = 0, \tilde{\omega} = -1$, which represent the exchange of energy from the dark energy into dark matter include a stable point $A (0,-\frac{3}{\gamma})$ which corresponds to dark energy dominant with diffusion effect. However, the second case $\omega = -1, \tilde{\omega} = 0$, which represent the exchange of energy from the dark matter into dark energy includes a stable point $C (0,0)$ which corresponds to dark energy dominant with no diffusion effect.

In this model we have chosen $\omega$ and $\tilde{\omega}$ being constants, whereas in general Lagrangian models $\omega$ and $\tilde{\omega}$ are varying in time. However we expect that $\omega$ and $\tilde{\omega}$ can be approximated by constants for large times. In the next sections we investigate more general dynamics on the basis of the action principle.

\section{A Lagrangian with Dynamical Space-time}
\subsection{Two Measures Theories} 
The Two Measure Theory implies other measure of integration in addition to the regular measure of integration in the action $ \sqrt{-g} $. The new measure is also a density and a total derivative. A simple example for constructing this measure is by introducing 4 scalar fields $ \varphi_{a} $, where $ a=1,2,3,4 $. The measure reads:
\begin{equation}
\Phi=\varepsilon^{\alpha\beta\gamma\delta}\varepsilon_{abcd}\partial_{\alpha}\varphi_{a}\partial_{\beta}\varphi_{b}\partial_{\gamma}\varphi_{c}\partial_{\delta}\varphi_{d} 
\end{equation}
A complete action involves both measures takes the form:
\begin{equation}\label{bTMT}
S=\int d^{4}x\Phi\mathcal{L}_{1}+\int d^{4}x\sqrt{-g}\mathcal{L}_{2}
\end{equation}
As a consequence of the variation with respect to the scalar fields $ \varphi_{a} $, under the assumption that $ \mathcal{L}_{1} $
and $ \mathcal{L}_{2}$ are independent of the scalar fields $\varphi_{a} $, we obtain that:
\begin{equation} \label{measure}
A_{a}^{\alpha}\partial_{\alpha}\mathcal{L}_{1}=0
\end{equation}
where $A_{a}^{\alpha}=\varepsilon^{\alpha\beta\gamma\delta}\varepsilon_{abcd}\partial_{\beta}\varphi_{b}\partial_{\gamma}\varphi_{c}\partial_{\delta}\varphi_{d} $. Since $ \det[A_{a}^{\alpha}]\sim\Phi^{3} $ as one easily see then that for $ \Phi\neq0 $, Eq. (\ref{measure}) implies that $\mathcal{L}_{1}=M=Const$. These kind of contributions have been considered in the Two Measures Theories which are of interest in connection with a unified model of dark energy and dark matter \cite{Guendelman:2012gg}.
\subsection{Dynamical time action}
The constraint on the term in the action $\mathcal{L}_2$ as in the Two Measure Theories (\ref{bTMT}) could be generalized to a covariant conservation of a stress energy momentum tensor $T_{\left(\chi\right)}^{\mu\nu}$ which coupled directly in the action \cite{Guendelman:2009ck}:
\begin{equation}\label{1}
\mathcal{S}= \int d^{4}x\sqrt{-g} \, \chi_{\mu;\nu}T_{\left(\chi\right)}^{\mu\nu}
\end{equation}
to a vector field $\chi_\mu$ with it's covariant derivatives $ \chi_{\mu;\nu}=\partial_{\nu}\chi_{\mu}-\Gamma_{\mu\nu}^{\lambda}\chi_{\lambda}$. From the variation with respect to the vector field $\chi_\mu$ gives a constraint on the conservation of the stress energy tensor $T_{\left(\chi\right)}^{\mu\nu}$. 
\begin{equation} 
\delta\chi_\mu: \nabla_\mu T_{\left(\chi\right)}^{\mu\nu}=0
\end{equation}
Similarly as the variation with respect to the scalar field $\varphi_a$ in the Lagrangian (\ref{bTMT}) yields $\partial_\alpha \mathcal{L}=0$. The correspondence between them is when $T^{\mu\nu}_{(\chi)}$ is taken to be as $T^{\mu\nu}_{(\chi)} = g^{\mu\nu} \mathcal{L}_m$. By introducing the term in the action (\ref{1}), we get:
\begin{equation}\label{CM}
\begin{split}
\int d^{4}x\,\sqrt{-g}\chi_{\mu;\nu}T_{\left(\chi\right)}^{\mu\nu} =\int d^{4}x\,\sqrt{-g}\chi^{\lambda}_{;\lambda} \mathcal{L}_m \\ = \int d^{4}x \,\partial_\mu (\sqrt{-g} \chi^\mu) \mathcal{L}_m = \int d^{4}x \, \Phi \mathcal{L}_m 
\end{split}
\end{equation}
Similarly to the variation (\ref{bTMT}), the variation with respect to the scalar field gives again $\partial_\mu\mathcal{L}_m=0$. For dynamical time theories, the variation with respect to the dynamical time vector field yields the same constraint.

The name Dynamical Time Theory (DTT) was considered due to the fact the energy density $T^{0}_{0} {\left(\chi\right)}$ is the canonically conjugated variable to the dynamical time $\chi^{0}$:
\begin{equation}\label{HMHM}
\pi_{\chi^0}=\frac{\partial\mathcal{L}}{\partial\dot{\chi^0}}=T^{0}_{0} (\chi):=\rho_{(\chi)}
\end{equation}
where $\rho_{(\chi)}$ is the energy density of the original stress energy tensor. 

\subsection{Dynamical time action with diffusive source}
In order to break the conservation of $T^{\mu\nu}_{(\chi)}$ as in the diffusion equation (Eq. \ref{diffusion}), the vector field $\chi_\mu$ should be coupled in a mass like term in the action:
\begin{equation}\label{nhd1}
\begin{split}
S_{(\chi,A)}=\int d^{4}x\sqrt{-g}\chi_{\mu;\nu}T_{\left(\chi\right)}^{\mu\nu} \\+ \frac{\kappa}{2}\int d^4x \sqrt{-g}(\chi_{\mu}+\partial_{\mu}A)^2 
\end{split}
\end{equation} 
where $A$ is a  scalar field different from $\phi$. From a variation with respect to the dynamical space time vector field $\chi_{\mu}$ we obtain:
\begin{equation} \label{nhd2}
\nabla_{\nu}T_{\left(\chi\right)}^{\mu\nu}=\kappa(\chi^{\mu}+\partial^{\mu}A)= f^\mu,
\end{equation} 
where the current source reads: $f^\mu=\kappa (\chi^{\mu}+\partial^{\mu}A)$. From the variation with respect to the new scalar $A$ a covariant conservation of the current  indeed emerges: 
\begin{equation}\label{nhd3}
\nabla_{\mu}f^\mu=\kappa\nabla_{\mu}(\chi^{\mu}+\partial^{\mu}A)=0
\end{equation} 
The stress energy tensor $T_{\left(\chi\right)}^{\mu\nu}$ is substantially different from stress energy tensor that we all know from Einstein equation which is defined as $\frac{8\pi G}{c^4}T^{\mu\nu}_{(G)}=R^{\mu\nu}-\frac{1}{2}g^{\mu\nu}R$. In this case, the stress energy momentum tensor $T^{\mu\nu}_{(\chi)}$ is a diffusive non conservative stress energy tensor. However, from a variation with respect to the metric, we get the conserved stress energy tensor as in Einstein equation:
\begin{equation}
T^{\mu\nu}_{(G)}=\frac{-2}{\sqrt{-g}}\frac{\delta(\sqrt{-g}\mathcal{L}_M)}{\delta g^{\mu\nu}}\quad , \quad \nabla_{\mu}T_{\left(G\right)}^{\mu\nu}=0 
\end{equation}
Using different expressions for $T^{\mu\nu}_{(\chi)}$ which depends on different variables, gives the conditions between the dynamical space time vector field $\chi_\mu$ and the other variables.
\subsection{Higher derivatives action}
A particular case of diffusive energy theories is obtained when $\sigma \to \infty$. In this case, the contribution of the current $f_\mu$ in the equations of motion goes to zero and  yields a constraint for the vector field being a gradient of the scalar:
\begin{equation}
f_\mu=\kappa(\chi_{\mu}+\partial_{\mu}A) =0 \quad \Rightarrow \quad \chi_{\mu}=-\partial_{\mu}A
\end{equation}
For the rest of the paper we use the notation $\chi$ for the scalar field which is coupled to the stress energy momentum tensor and not $A$ due to earlier publications. The theory (\ref{nhd1}) is reduced to a theory with higher derivatives:
\begin{figure*}[t!]
 	\centering
\includegraphics[width=0.9\textwidth]{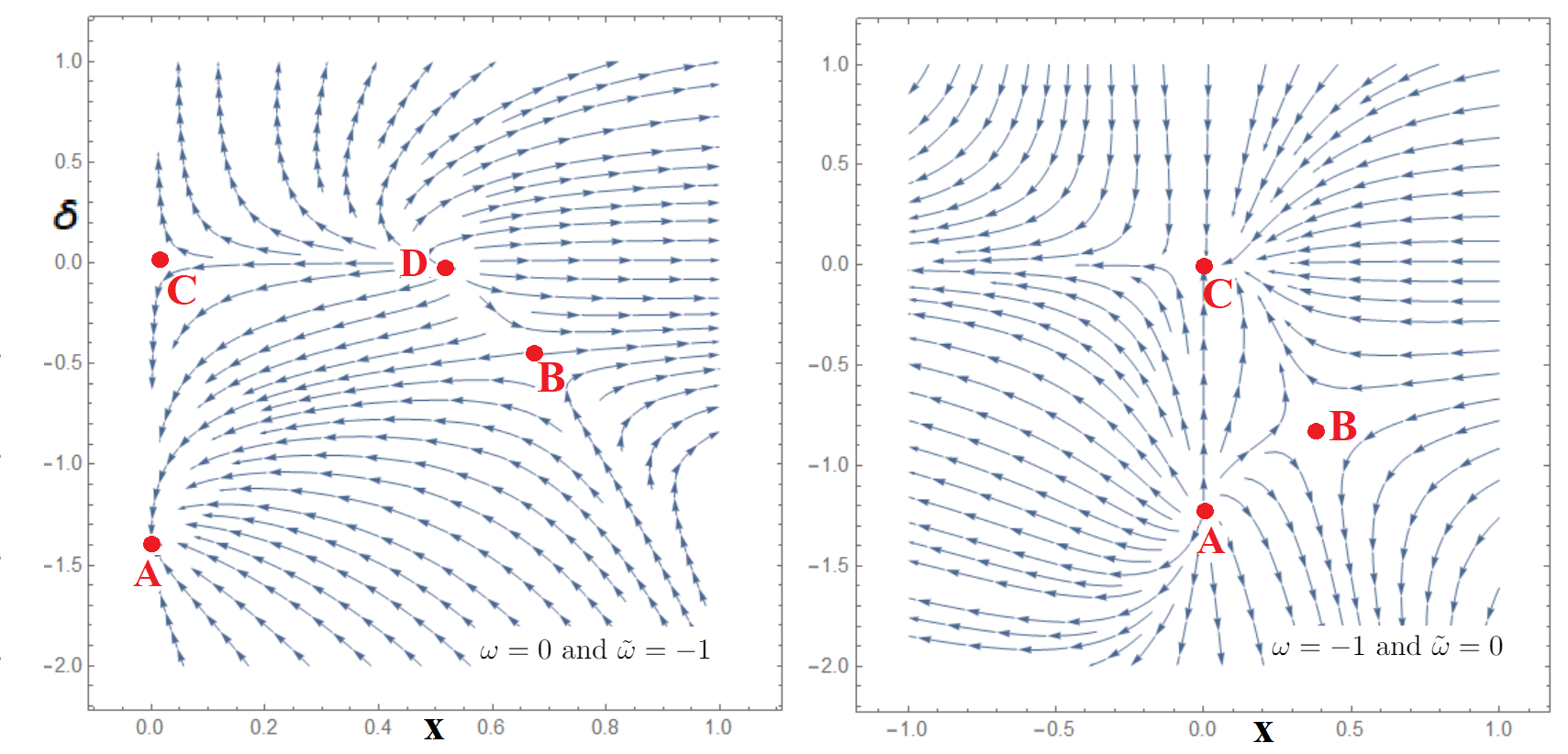}
\caption{The phase portrait for the dynamical system method. In the left panel the $\tilde{\omega} = -1$ refers to dark energy and in the right panel the $\omega = -1$ refers to dark energy}
 	\label{fig}
 \end{figure*}
\begin{equation}\label{action2}
\mathcal{S}= - \int d^{4}x\sqrt{-g} \, \nabla_\mu \nabla_\nu \chi \, \,   T_{\left(\chi\right)}^{\mu\nu}
\end{equation}
The variation with respect to the scalar $A$ gives $\nabla_\mu \nabla_\nu T^{\mu\nu}_{(\chi)}=0$ which  corresponds to the variations (\ref{nhd2}) - (\ref{nhd3}). In the following paper we use the reduced theory with higher derivative in the action. 
\section{Scalar field Gravity with Diffusive behavior}
\subsection{Dynamical time action with diffusive source}
In this section we consider the following action:
\begin{equation}
\mathcal{L} = \frac{1}{2}\mathcal{R} + \chi_{,\mu;\nu}T^{\mu\nu}_{(\chi)} -\frac{1}{2}\phi^{,\mu}\phi_{,\mu} - V(\phi)
\end{equation}
which contains a scalar field with potential $V(\phi)$. The stress energy momentum tensor $T^{\mu\nu}_{(\chi)}$ is chosen to be:   
\begin{equation}
T^{\mu\nu}_{(\chi)} = -\frac{\lambda_1}{2} \phi^{,\mu}\phi^{,\nu} - \frac{\lambda_2}{2} g^{\mu\nu} (\phi_{,\alpha}\phi^{,\alpha}) + g^{\mu\nu} U(\phi)
\end{equation} 
where $\lambda_1$ and $\lambda_2$ are arbitrary constants, and $U(\phi)$ is a another potential. In such a case the density and pressure resulting from $T^{\mu\nu}_{(\chi)}$ are:
\begin{equation}
\rho_{(\chi)} = (\lambda_1+\lambda_2)\frac{\dot{\phi}^2}{2} + U(\phi),
\end{equation}
\begin{equation}
p_{(\chi)} = -\lambda_2\frac{\dot{\phi}^2}{2} - U(\phi)
\end{equation}
Notice that the starting point was the case of two fluids. But here we discuss about single fluid with a Lagrangian involving two different measures: where the modified measure is generalized by using the dynamical space time vector field $\chi_\mu$.  

There are three independent sets of equations of motions: $\chi$, $\phi$ and the metric $g_{\mu\nu}$.
The variation with respect to the field $\chi$ yields:
\begin{equation}\label{chi}
\nabla_{\mu} \nabla_{\nu}  T^{\mu\nu}_{(\chi)} = 0
\end{equation}
The variation with respect to the field $\phi$ gives a non-conserved current $j^{\mu}$:
\begin{equation}\label{phi1}
j^{\mu} = \frac{\lambda_1}{2}(\chi^{,\mu;\nu}+\chi^{,\nu;\mu})\phi_{,\nu} + (1+ \lambda_2 \Box \chi) \phi^{,\mu},
\end{equation}
with the the non conservation law:
\begin{equation}\label{phi2}
 \nabla_{\mu} j^{\mu} = V'(\phi) - \Box \chi \, U'(\phi) 
\end{equation}
The Einstein equations derived from the variation with respect to the metric take the form: 
\begin{equation}\label{Einstein}
\begin{split}
G^{\mu\nu} = g^{\mu\nu} \left( -\chi_{,\alpha;\beta}T^{\alpha\beta}_{(\chi)} +\frac{1}{2}\phi^{,\alpha}\phi_{,\alpha} + V(\phi) \right)\\- \phi^{,\mu}\phi^{,\nu} + \chi_{,\alpha;\beta} \frac{\partial T^{\alpha\beta}_{(\chi)}}{g_{\mu\nu}} \\ + \nabla_\lambda \left( \chi^{,\mu} T^{\nu\lambda}_{(\chi)}+\chi^{,\nu} T^{\mu\lambda}_{(\chi)} - \chi^{,\lambda} T^{\mu\nu}_{(\chi)} \right)
\end{split}
\end{equation}
where the derivative of the energy momentum tensor $T^{\mu\nu}_{(\chi)}$ with respect to $g_{\mu\nu}$  yields:
\begin{equation*} 
\begin{split}
\chi_{,\alpha;\beta} \frac{\partial T^{\alpha\beta}}{\partial g_{\mu\nu}} = -\frac{\lambda_1}{2} \chi^{(,\mu}\phi^{,\nu)} \Box \phi + (\frac{\lambda_1}{2}+ \lambda_2) \phi^{,\mu}\phi^{,\nu} \Box \chi \\ +\frac{\lambda_1}{2} \chi^{,\gamma;\mu} \phi^{,\nu} \phi_{,\gamma} - \lambda_2 \phi^{,\mu;\lambda} \chi^{,\nu}\phi_{,\lambda} - \lambda_2 \chi^{,\mu} \phi^{,\gamma;\nu}\phi_{,\nu} \\+ \frac{\lambda_1}{2}\phi^{,\mu} \chi^{,\gamma;\nu}\phi_{,\gamma} - \frac{\lambda_1}{2} \chi^{(,\nu} \phi^{,\mu);\gamma} \phi_{,\gamma} \\+ \frac{\lambda_1}{2} \pi^{(,\nu} \phi^{,\mu);\gamma} \chi_{,\gamma} + \chi^{(,\mu} \phi^{,\nu)} U'(\phi)
\end{split}
\end{equation*}
The expression in the right hand side of Eq. (\ref{Einstein}) is the total energy momentum tensor.
\subsection{Cosmological solution}
For the solution we assume homogeneity and isotropy, therefore we solve our theory with a FLRW metric:
\begin{equation}
ds^2=-dt^2+a(t)^2(\frac{dr^2}{1-K r^2}+r^2 d\Omega^2)
\end{equation}
According to this ansatz the scalar fields are solely functions of time.

Integrating Eq. ($\ref{chi}$) once, we express it in the form:
\begin{equation}\label{chisol}
(\lambda_1+\lambda_2)\dot{\phi}\ddot{\phi} + U'(\phi)\dot{\phi} + 3 H \lambda_1\dot{\phi}^2 = \frac{\sigma}{a^3}
\end{equation}
where $\sigma$ is an integration constant.

For dark energy dynamics we can assume that $U(\phi) = \textbf{const}$. Then the solution for Eq. (\ref{chisol}) is:
\begin{equation}\label{chisol2}
\dot{\phi}^2  = \dot{\phi}_{(0)}^2 a^{-\frac{3\lambda_1}{\lambda_1+\lambda_2}}+ \frac{\sigma}{\lambda_1+\lambda_2} a^{-\frac{3\lambda_1}{\lambda_1+\lambda_2}} \int_{0}^{t}ds a^{-\frac{3\lambda_2}{\lambda_1+\lambda_2}} 
\end{equation}
In addition for the same theoretical reason we assume that $V(\phi) = \textbf{Const}$. Then the current conservation law ($\ref{phi2}$) has the solution:
\begin{equation}\label{chi3}
(\frac{\lambda_1}{2}-\lambda_2) \ddot{\chi} + (1 - 3H \dot{\chi})\lambda_2  = \frac{\tilde{\sigma}}{\dot{\phi} a^3}
\end{equation}
where $\tilde{\sigma}$ is another integration constant. Now from the stress energy momentum tensor the total energy density term is:
\begin{equation}\label{density}
\begin{split}
\rho = \frac{3}{2} H (\lambda_1-2 \lambda_2) \dot{\chi} \dot{\phi}^2\\+\frac{1}{2} \dot{\phi}^2 \left(1-2 (\lambda_1+\lambda_2) \ddot{\chi}\right)+\dot\chi \dot\phi \left((\lambda_1+\lambda_2) \ddot\phi\right)+V,
\end{split}
\end{equation}
and the total pressure is:
\begin{equation}
p = \frac{1}{2} \dot\phi^2 -\frac{1}{2}\lambda_1 \ddot\chi \dot\phi^2+\lambda_2 \dot\chi \dot\phi \ddot\phi-V.
\end{equation}
\section{Asymptotic solutions}
We aren't able to find the exact solutions for the Einstein equation (\ref{Einstein}) together with the equations for the scalar fields $\chi$ (Eq. \ref{chisol}) and $\phi$ (Eq. \ref{chi3}). So we are looking for asymptotic solutions.
\subsection{A Power Law Solution}
We assume a power law solution for a large time:
\begin{equation}
a \sim t^{\alpha}
\end{equation}
Then from Eq. (\ref{chisol}) the solution for the scalar field $\phi$ derivative is:
\begin{equation}\label{phipower}
\dot{\phi} = \sqrt{\frac{2\sigma}{3 \alpha (\lambda_1-\lambda_2)+\lambda_1+\lambda_2}}\, t^{\frac{1}{2}-\frac{3 \alpha }{2}}
\end{equation}
where $\phi_0$ is an arbitrary integration constant. 

The solution for the scalar field $\chi$ is:
 \begin{equation}\label{chipower}
 \dot{\chi} = C t
 \end{equation}
with the constant:
\begin{equation}
C = \frac{2\lambda_2}{-6 \alpha  \lambda_2+\lambda_1-2 \lambda_2}
\end{equation}
By inserting the solutions (\ref{phipower}) and (\ref{chipower}) into Einstein equation we obtain:
\begin{equation}
\rho = \frac{\alpha_1}{a^3} + \frac{\alpha_2 t}{a^3}+V
\end{equation}
where the constants are:
\begin{equation}
\alpha_1 = \frac{18 \alpha ^2 \lambda_2 (2 \lambda_2-\lambda_1)}{2 (\lambda_1-2\lambda_2 (3 \alpha  +1))}
\end{equation}
\begin{equation}
\alpha_2 = \frac{(6 \alpha +2) \lambda_1 \lambda_2+2 (3 \alpha +1) (\lambda_2-1)\lambda_2+ \lambda_1}{2 (\lambda_1-2 \lambda_2 (3 \alpha +1))}
\end{equation}
We get an asymptotic solution if the potential $V = 0$ and the power of the scale factor is one:
\begin{equation}
a \sim t
\end{equation}
This solutions is same as the one obtained in the model of Einstein equation with relativistic diffusion exchange of energy \cite{Haba:2016swv}.  
\subsection{Exponential Solution}
We insert the exponential solution $a(t) \sim e^{H_0 t}$ in Eq. (\ref{chisol2}). Then we get: 
\begin{equation}\label{scalarExp}
\dot{\phi}^2 = \dot{\phi}_0^2 a^{-\frac{3\lambda_1}{\lambda_1+\lambda_2}} - \sigma_0 H_0 \frac{\lambda_1+\lambda_2}{3\lambda_2}\frac{1}{a^3}
\end{equation}
if we impose $\frac{3\lambda_1}{\lambda_1+\lambda_2} > 0$. Then from Eq. (\ref{chi3}) we obtain the asymptotic solution: 
\begin{equation}
\dot{\chi} = \frac{1}{3H_0} + \mathcal{O}(\frac{1}{a^3})
\end{equation}
With those solutions the density is given by:
\begin{equation}
\begin{split}
\rho = H_0 (3 \lambda_2-1) \sigma\frac{\lambda_1+\lambda_2}{6\lambda_2}\frac{1}{a^3} + V \\ + \frac{1}{2} \dot{\phi}_0^2 (1 - 2\lambda_2) a^{-\frac{3\lambda_1}{\lambda_1+\lambda_2}}
\end{split}
\end{equation}
This particular solution corresponds to a slowly varying dark energy ($V + \frac{1}{2} \dot{\phi}_0^2 (1 - 2\lambda_2) a^{-\frac{3\lambda_1}{\lambda_1+\lambda_2}}$) approaching a constant value $V$, for $\lambda_1$ and $\lambda_2$ being positive, and $\lambda_1 \ll \lambda_2$. In the case of negative $\lambda_1$ but still $|\lambda_1| \ll |\lambda_2|$, we get slowly growing vacuum energy, which corresponds to an asymptotically super accelerating universe.
\section{$\lambda_2 = 0$ Case}
Solution (\ref{scalarExp}) does not make sense for $\lambda_{2}=0$. Therefore this case should be treated separately. This special choice of the energy momentum has been explored by Gao, Kunz, Liddle and Parkison as a unification of dark energy and dark matter \cite{Gao:2009me} without using a  lagrangian formulation. These authors proposed
as a unification of dark energy and dark matter :
\begin{equation}
T^{\mu\nu}_{(\chi)} = -\frac{\lambda_1}{2}\phi^{,\mu}\phi^{,\nu} + g^{\mu\nu} U(\phi)
\end{equation}
as the right hand side of Einstein tensor. The action that produces asymptotically the same model using dynamical time theories was obtained in Ref. \cite{Benisty:2018qed}. Here we explore the asymptotic solution with diffusive behavior. Under the assumption that all of the potentials are constant Eq. (\ref{chisol}) has the solution:
\begin{equation}
\dot{\phi}^2  = \frac{\dot{\phi}_{(0)}^2}{a^3}+ \frac{\sigma}{\lambda_1} \frac{t}{a^3} 
\end{equation}
Then, the integral of Eq. (\ref{chi3}) is :
\begin{equation}
\dot{\chi} (t) =\dot{\chi} (0) -\frac{2}{\lambda_1}t + \int \,dt \frac{2\tilde{\sigma}}{\lambda_1 \dot{\phi} a^3} 
\end{equation}
with the asymptotic behavior:
\begin{equation}
\dot{\chi} (t \rightarrow \infty) \rightarrow -\frac{2t}{\lambda_1}.
\end{equation}
Notice that this asymptotic behavior is essentially different from the previous cases. Then the total density reads:
\begin{equation}\label{densityLiddle}
\rho = V + \frac{\alpha_1}{a^3} + \frac{\alpha_2}{a^{4.5}},
\end{equation}
where the coefficients are:
\begin{equation}
\alpha_1 =  \frac{5 \dot{\phi}_0^2 \lambda_1+\lambda_1 \sigma \chi_0 + 3 \sigma t}{2\lambda_1}
\end{equation}
\begin{equation}\label{alpha2}
\alpha_2=-\frac{2 \tilde{\sigma}}{3 \dot{\phi}_0 H_0 \lambda_1} \left(3 \dot{\phi}_0^2 H_0 \lambda_1+3 H_0 \sigma  t+\sigma \right)
\end{equation}
Additional symmetry for this case is obtained: 
\begin{equation}
\chi \rightarrow \chi + ct
\end{equation}
or in terms of the dynamical time ($\chi^0 \Leftrightarrow \dot{\chi}$)
\begin{equation}\label{symet}
\chi^0 \rightarrow \chi^0 + c
\end{equation}
In the previous cases $\dot{\chi}$ is asymptotically a constant, equal to $\frac{1}{3H_0}$. In the special case of $\lambda_2 = 0 $ there cannot be any particular choice for asymptotic value of $\chi$, because the symmetry will change it to any other arbitrary constant. One can calculate the conserved quantity associated with the symmetry (\ref{symet}) and it is the analogous of particle number.

A remarkable result is the correspondence between the solution (\ref{densityLiddle}) and the solutions for the DM-DE interaction system from Sec (2). For $\tilde{\omega} = 0$ the dust density equation yields:
\begin{equation}
\partial_t \rho_{dust} + 3 H \rho_{dust} = \frac{\gamma^2}{a^3}, 
\end{equation}
with the solution:
\begin{equation}
\rho_{dust} = \frac{C_1}{a^3} + \frac{\gamma^2 t}{a^3}.
\end{equation}
where $C_1$ is an integration constant. For interacting dark energy, that satisfies $\omega = -1$, the energy density reads:
\begin{equation}
\partial_t \rho_{\Lambda} = -\frac{\gamma^2}{a^3},
\end{equation}
whereas for $\rho_{\Lambda}$
\begin{equation}
\rho_{\Lambda} = C_2 - \gamma^2 \int{\frac{dt}{a^3}}.
\end{equation}
The $C_2$ is another integration constant. Asymptotically, the total density gives:
\begin{equation}
\rho = \frac{C_1+\gamma^2 t}{a^3} + C_2 + \mathcal{O}(\frac{1}{a^6}),
\end{equation}
which corresponds to the density (\ref{densityLiddle}), and the last term $\frac{\alpha_2}{a^{4.5}}$ becomes negligible. Hence, the integration constants equals to the integration constants from the Lagrangian case:
\begin{equation}\label{c1}
C_1 =  \frac{5 \dot{\phi}_0^2+\sigma \chi_0 }{2}
\end{equation}
\begin{equation}\label{c2}
\gamma^2 = \frac{3 \sigma}{2\lambda_1}, \quad C_2 = V
\end{equation}
This correspondence does not hold for the whole history of the universe, however asymptotically the models (our Lagrangian model and the previously studied non Lagrangian models) fit each other
for the case $\lambda_2 =0$ and approach $\Lambda$CDM for late times. Of course that the solutions will have to be studied and this will be a main goal for further investigations.

One can see that both models with exactly the same homogeneous solution where $\tilde{\sigma} = 0$. In this case $alpha_2 = 0$ (see Eq. \ref{alpha2}) and the corresponding relations between the constants of the models present in Eq. (\ref{c1} - \ref{c2}).

In order to assess the viability of the model, let us see how some physical quantities change versus the red-shift (z) for both models. The connection between the cosmic time derivative and the red-shift derivative reads:
 \begin{figure*}[t!]
 	\centering
\includegraphics[width=1\textwidth]{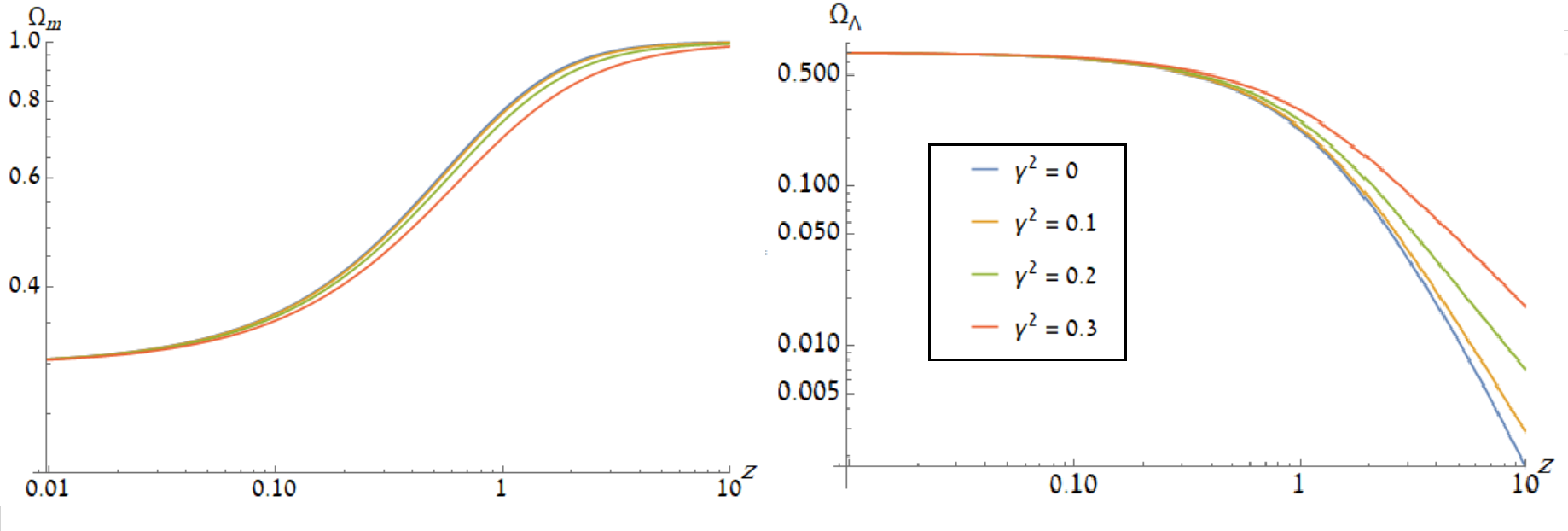}
\caption{The numerical solution of the partial densities of the dark energy and dark matter components, for different values of the coupling $\gamma^2$ (which is corresponding to the diffusion constant $\sigma$).}
 	\label{fig2}
 \end{figure*}
\begin{equation}
\frac{d}{dt} = -H(z) (z+1) \frac{d}{dz}
\end{equation}
which is obtained from the dependence of scale factor on the red-shift $a = \frac{1}{z+1}$. The numerical solution of the partial densities for the simplest case appear in Fig. (\ref{fig2}). Even this simple case describes a diffusive interaction between dark energy dark matter from an action principle. However, the presence of the coupling constant $\tilde{\sigma}$ yields to additional part ($\sim a^{-4.5}$) which could resolve the singularity problem as discussed in Ref. \cite{Benisty:2018qed}. But in any case - all the solutions approach $\Lambda CDM$ model for the late universe. 

\section{Conclusions}
We have extended the results of our earlier papers concerning the DM-DE  interaction in the context of two measures models and the dynamical time theories.
The extension consists in a general choice of the conserved non-canonical energy-momentum tensor. The energy momentum tensor is more general than the one
proposed by Gao, Kunz, Liddle and Parkinson \cite{Gao:2009me} as well as the Dark Energy Dark Matter unification obtained in the Two Measures limit, which corresponds to the case where the conserved non-canonical energy-momentum tensor is proportional to the metric tensor \cite{Guendelman:2015rea},\cite{Guendelman:2012gg}. 

The constants $\lambda_1$ and $\lambda_2$ parametrize the more general choice considered here. $\lambda_2 =0 $ corresponds to the case considered by Gao, Kunz, Liddle and Parkinson in their non-Lagrangian formalism, In our Lagrangian formulation, for this type of energy momentum tensor, as additional shift symmetry  for the dynamical time appears and at the same time the dynamical time behaves asymptotically as the cosmic time. Diffusive type is obtained when the dynamical space time vector is taken to be the gradient of a scalar, then instead of a conservation law of the energy momentum introduced in the action, we obtain a non conservation of this energy momentum tensor of the diffusive type, which leads then to an interacting DE/DM scenario. This formulation of DE-DM have a direct correspondence with the behavior of non Lagrangian formulations of DE/DM interactions only in the case $\lambda_2 =0 $. In the other cases, the asymptotic behavior is different and in particular the dynamical time does not behave as cosmic time asymptotically, in fact as the cosmic time increases, the dynamical time approaches the finite value $\frac{1}{3H}$ in an asymptotically de-Sitter space. In all cases we do not need to introduce the dark matter in the initial Lagrangian, it appears dynamically.  As a result of the dynamic evolution in our model we obtain an asymptotically $\Lambda$CDM solution.

\acknowledgments
This article is supported by COST Action CA15117 "Cosmology and Astrophysics Network for Theoretical Advances and Training Action" (CANTATA) of the COST (European Cooperation in Science and Technology).


\begin{thebibliography}{99}
\bibitem{Cardone:2004sq} 
  V.~F.~Cardone, A.~Troisi and S.~Capozziello,
  %``Unified dark energy models: A Phenomenological approach,''
  Phys.\ Rev.\ D {\bf 69}, 083517 (2004)
  doi:10.1103/PhysRevD.69.083517
  [astro-ph/0402228].
  %%CITATION = doi:10.1103/PhysRevD.69.083517;%%
  \bibitem{Capozziello:2017buj} 
  S.~Capozziello, R.~D'Agostino and O.~Luongo,
  %``Cosmic acceleration from a single fluid description,''
  Phys.\ Dark Univ.\  {\bf 20}, 1 (2018)
  doi:10.1016/j.dark.2018.02.002
  [arXiv:1712.04317 [gr-qc]].
  %%CITATION = doi:10.1016/j.dark.2018.02.002;%%
  %9 citations counted in INSPIRE as of 22 Feb 2019
  \bibitem{Capozziello:2018mds} 
  S.~Capozziello, R.~D'Agostino, R.~Giambò and O.~Luongo,
  %``Effective field description of the Anton-Schmidt cosmic fluid,''
  Phys.\ Rev.\ D {\bf 99}, no. 2, 023532 (2019)
  doi:10.1103/PhysRevD.99.023532
  [arXiv:1810.05844 [gr-qc]].
  %%CITATION = doi:10.1103/PhysRevD.99.023532;%%
  %3 citations counted in INSPIRE as of 22 Feb 2019
\bibitem{Dvali:2003rk} 
  G.~Dvali and M.~S.~Turner,
  %``Dark energy as a modification of the Friedmann equation,''
  astro-ph/0301510.
  %%CITATION = ASTRO-PH/0301510;%%
  %203 citations counted in INSPIRE as of 11 Feb 2018
\bibitem{Basilakos:2009ah} 
  S.~Basilakos,
  %``Cosmological implications and structure formation from a time varying vacuum,''
  Mon.\ Not.\ Roy.\ Astron.\ Soc.\  {\bf 395}, 2347 (2009)
  doi:10.1111/j.1365-2966.2009.14713.x
  [arXiv:0903.0452 [astro-ph.CO]].
  %%CITATION = doi:10.1111/j.1365-2966.2009.14713.x;%%
  %17 citations counted in INSPIRE as of 11 Feb 2018
\bibitem{Arevalo:2016epc} 
  F.~Arevalo, A.~Cid and J.~Moya,
  %``AIC and BIC for cosmological interacting scenarios,''
  Eur.\ Phys.\ J.\ C {\bf 77}, no. 8, 565 (2017)
  doi:10.1140/epjc/s10052-017-5128-7
  [arXiv:1610.09330 [astro-ph.CO]].
  %%CITATION = doi:10.1140/epjc/s10052-017-5128-7;%%
  %7 citations counted in INSPIRE as of 05 Dec 2018
   \bibitem{Scherrer:2004au} 
  R.~J.~Scherrer,
  %``Purely kinetic k-essence as unified dark matter,''
  Phys.\ Rev.\ Lett.\  {\bf 93}, 011301 (2004)
  doi:10.1103/PhysRevLett.93.011301
  [astro-ph/0402316].
  %%CITATION = doi:10.1103/PhysRevLett.93.011301;%%
  %335 citations counted in INSPIRE as of 11 Feb 2018
    \bibitem{Fadragas:2013ina}
  C.~R.~Fadragas, G.~Leon and E.~N.~Saridakis,
  %``Dynamical analysis of anisotropic scalar-field cosmologies for a wide range of potentials,''
  Class.\ Quant.\ Grav.\  {\bf 31} (2014) 075018
  doi:10.1088/0264-9381/31/7/075018
  [arXiv:1308.1658 [gr-qc]].
  %%CITATION = doi:10.1088/0264-9381/31/7/075018;%%
  %35 citations counted in INSPIRE as of 06 Dec 2018
  \bibitem{Ratra:1987rm} 
  B.~Ratra and P.~J.~E.~Peebles,
  %``Cosmological Consequences of a Rolling Homogeneous Scalar Field,''
  Phys.\ Rev.\ D {\bf 37}, 3406 (1988).
  doi:10.1103/PhysRevD.37.3406
  %%CITATION = doi:10.1103/PhysRevD.37.3406;%%
  %3318 citations counted in INSPIRE as of 17 Nov 2018
\bibitem{Luongo:2018lgy} 
  O.~Luongo and M.~Muccino,
  %``Speeding up the universe using dust with pressure,''
  Phys.\ Rev.\ D {\bf 98}, no. 10, 103520 (2018)
  doi:10.1103/PhysRevD.98.103520
  [arXiv:1807.00180 [gr-qc]].
  %%CITATION = doi:10.1103/PhysRevD.98.103520;%%
  %8 citations counted in INSPIRE as of 22 Feb 2019
\bibitem{Arbey:2006it} 
  A.~Arbey,
  %``Dark fluid: A Complex scalar field to unify dark energy and dark matter,''
  Phys.\ Rev.\ D {\bf 74}, 043516 (2006)
  doi:10.1103/PhysRevD.74.043516
  [astro-ph/0601274].
    \bibitem{Chen:2008ft} 
  X.~M.~Chen, Y.~G.~Gong and E.~N.~Saridakis,
  %``Phase-space analysis of interacting phantom cosmology,''
  JCAP {\bf 0904}, 001 (2009)
  doi:10.1088/1475-7516/2009/04/001
  [arXiv:0812.1117 [gr-qc]].
  %%CITATION = doi:10.1088/1475-7516/2009/04/001;%%
  %183 citations counted in INSPIRE as of 06 Dec 2018
  \bibitem{Leon:2009rc}
  G.~Leon and E.~N.~Saridakis,
  %``Phase-space analysis of Horava-Lifshitz cosmology,''
  JCAP {\bf 0911} (2009) 006
  doi:10.1088/1475-7516/2009/11/006
  [arXiv:0909.3571 [hep-th]].
  %%CITATION = doi:10.1088/1475-7516/2009/11/006;%%
  %103 citations counted in INSPIRE as of 06 Dec 2018
  \bibitem{Leon:2010pu} 
  G.~Leon and E.~N.~Saridakis,
  %``Dynamics of the anisotropic Kantowsky-Sachs geometries in $R^n$ gravity,''
  Class.\ Quant.\ Grav.\  {\bf 28}, 065008 (2011)
  doi:10.1088/0264-9381/28/6/065008
  [arXiv:1007.3956 [gr-qc]].
  %%CITATION = doi:10.1088/0264-9381/28/6/065008;%%
  %52 citations counted in INSPIRE as of 06 Dec 2018
  \bibitem{Leon:2012mt} 
  G.~Leon and E.~N.~Saridakis,
  %``Dynamical analysis of generalized Galileon cosmology,''
  JCAP {\bf 1303}, 025 (2013)
  doi:10.1088/1475-7516/2013/03/025
  [arXiv:1211.3088 [astro-ph.CO]].
  %%CITATION = doi:10.1088/1475-7516/2013/03/025;%%
  %69 citations counted in INSPIRE as of 06 Dec 2018
  \bibitem{Leon:2014yua}
  G.~Leon and E.~N.~Saridakis,
  %``Dynamical behavior in mimetic F(R) gravity,''
  JCAP {\bf 1504} (2015) no.04,  031
  doi:10.1088/1475-7516/2015/04/031
  [arXiv:1501.00488 [gr-qc]].
  %%CITATION = doi:10.1088/1475-7516/2015/04/031;%%
  %67 citations counted in INSPIRE as of 06 Dec 2018
  \bibitem{Leon:2013qh} 
  G.~Leon, J.~Saavedra and E.~N.~Saridakis,
  %``Cosmological behavior in extended nonlinear massive gravity,''
  Class.\ Quant.\ Grav.\  {\bf 30}, 135001 (2013)
  doi:10.1088/0264-9381/30/13/135001
  [arXiv:1301.7419 [astro-ph.CO]].
  %%CITATION = doi:10.1088/0264-9381/30/13/135001;%%
  %64 citations counted in INSPIRE as of 06 Dec 2018
  \bibitem{Leon:2012mt} 
  G.~Leon and E.~N.~Saridakis,
  %``Dynamical analysis of generalized Galileon cosmology,''
  JCAP {\bf 1303}, 025 (2013)
  doi:10.1088/1475-7516/2013/03/025
  [arXiv:1211.3088 [astro-ph.CO]].
  %%CITATION = doi:10.1088/1475-7516/2013/03/025;%%
  %69 citations counted in INSPIRE as of 06 Dec 2018
  \bibitem{Kofinas:2014aka}
  G.~Kofinas, G.~Leon and E.~N.~Saridakis,
  %``Dynamical behavior in $f(T,T_G)$ cosmology,''
  Class.\ Quant.\ Grav.\  {\bf 31} (2014) 175011
  doi:10.1088/0264-9381/31/17/175011
  [arXiv:1404.7100 [gr-qc]].
  %%CITATION = doi:10.1088/0264-9381/31/17/175011;%%
  %75 citations counted in INSPIRE as of 06 Dec 2018
  \bibitem{Skugoreva:2014ena}
  M.~A.~Skugoreva, E.~N.~Saridakis and A.~V.~Toporensky,
  %``Dynamical features of scalar-torsion theories,''
  Phys.\ Rev.\ D {\bf 91} (2015) 044023
  doi:10.1103/PhysRevD.91.044023
  [arXiv:1412.1502 [gr-qc]].
  %%CITATION = doi:10.1103/PhysRevD.91.044023;%%
  %29 citations counted in INSPIRE as of 06 Dec 2018
\bibitem{Guendelman:2015jii} 
  E.~Guendelman, E.~Nissimov and S.~Pacheva,
  %``Unified Dark Energy and Dust Dark Matter Dual to Quadratic Purely Kinetic K-Essence,''
    Eur.\ Phys.\ J.\ C {\bf 76}, no. 2, 90 (2016)
  doi:10.1140/epjc/s10052-016-3938-7
  [arXiv:1511.07071 [gr-qc]].
  %%CITATION = doi:10.1140/epjc/s10052-016-3938-7;%%
  %23 citations counted in INSPIRE as of 11 Feb 2018
\bibitem{Guendelman:2012gg} 
    E.~Guendelman, D.~Singleton and N.~Yongram,
  %``A two measure model of dark energy and dark matter,''
  JCAP {\bf 1211}, 044 (2012)
  doi:10.1088/1475-7516/2012/11/044
  [arXiv:1205.1056 [gr-qc]].
  %%CITATION = doi:10.1088/1475-7516/2012/11/044;%%
  %26 citations counted in INSPIRE as of 11 Feb 2018
\bibitem{Guendelman:2015rea} 
  E.~Guendelman, E.~Nissimov and S.~Pacheva,
  %``Dark Energy and Dark Matter From Hidden Symmetry of Gravity Model with a Non-Riemannian Volume Form,''
  Eur.\ Phys.\ J.\ C {\bf 75}, no. 10, 472 (2015)
  doi:10.1140/epjc/s10052-015-3699-8
  [arXiv:1508.02008 [gr-qc]].
  %%CITATION = doi:10.1140/epjc/s10052-015-3699-8;%%
  %21 citations counted in INSPIRE as of 11 Feb 2018
\bibitem{Ansoldi:2012pi} 
  S.~Ansoldi and E.~I.~Guendelman,
  %``Unified Dark Energy-Dark Matter model with Inverse Quintessence,''
  JCAP {\bf 1305}, 036 (2013)
  doi:10.1088/1475-7516/2013/05/036
  [arXiv:1209.4758 [gr-qc]].
  %%CITATION = doi:10.1088/1475-7516/2013/05/036;%%
  %11 citations counted in INSPIRE as of 11 Feb 2018
\bibitem{Guendelman:2016kwj} 
  E.~Guendelman, E.~Nissimov and S.~Pacheva,
  %``Quintessential Inflation, Unified Dark Energy and Dark Matter, and Higgs Mechanism,''
  Bulg.\ J.\ Phys.\  {\bf 44}, 15 (2017)
  [arXiv:1609.06915 [gr-qc]].
  %%CITATION = ARXIV:1609.06915;%%
  %8 citations counted in INSPIRE as of 11 Feb 2018
  \bibitem{Koutsoumbas:2017fxp} 
  G.~Koutsoumbas, K.~Ntrekis, E.~Papantonopoulos and E.~N.~Saridakis,
  %``Unification of Dark Matter - Dark Energy in Generalized Galileon Theories,''
  JCAP {\bf 1802}, no. 02, 003 (2018)
  doi:10.1088/1475-7516/2018/02/003
  [arXiv:1704.08640 [gr-qc]].
  %%CITATION = doi:10.1088/1475-7516/2018/02/003;%%
  %2 citations counted in INSPIRE as of 16 Apr 2018
   \bibitem{Benisty:2017eqh} 
  D.~Benisty and E.~I.~Guendelman,
  %``Interacting Diffusive Unified Dark Energy and Dark Matter from Scalar Fields,''
  Eur.\ Phys.\ J.\ C {\bf 77}, no. 6, 396 (2017)
  doi:10.1140/epjc/s10052-017-4939-x
  [arXiv:1701.08667 [gr-qc]].
  %%CITATION = doi:10.1140/epjc/s10052-017-4939-x;%%
  %11 citations counted in INSPIRE as of 11 Feb 2018
  \bibitem{Benisty:2017rbw}
  D.~Benisty and E.~I.~Guendelman,
  %``Unified DE–DM with diffusive interactions scenario from scalar fields,''
  Int.\ J.\ Mod.\ Phys.\ D {\bf 26} (2017) no.12,  1743021.
  doi:10.1142/S0218271817430210
  %%CITATION = doi:10.1142/S0218271817430210;%%
\bibitem{Calogero:2013zba} 
  S.~Calogero and H.~Velten,
  %``Cosmology with matter diffusion,''
  JCAP {\bf 1311}, 025 (2013)
  doi:10.1088/1475-7516/2013/11/025
  [arXiv:1308.3393 [astro-ph.CO]].
  %%CITATION = doi:10.1088/1475-7516/2013/11/025;%%
  %11 citations counted in INSPIRE as of 22 Feb 2019
  \bibitem{Guendelman:2012gg} 
  E.~Guendelman, D.~Singleton and N.~Yongram,
  %``A two measure model of dark energy and dark matter,''
  JCAP {\bf 1211}, 044 (2012)
  doi:10.1088/1475-7516/2012/11/044
  [arXiv:1205.1056 [gr-qc]].
  %%CITATION = doi:10.1088/1475-7516/2012/11/044;%%
  %30 citations counted in INSPIRE as of 06 Dec 2018
  \bibitem{Guendelman:2009ck} 
  E.~I.~Guendelman,
  %``Gravitational Theory with a Dynamical Time,''
  Int.\ J.\ Mod.\ Phys.\ A {\bf 25}, 4081 (2010)
  doi:10.1142/S0217751X10050317
  [arXiv:0911.0178 [gr-qc]].
  %%CITATION = doi:10.1142/S0217751X10050317;%%
  %18 citations counted in INSPIRE as of 06 Dec 2018
    \bibitem{Haba:2016swv} 
  Z.~Haba, A.~Stachowski and M.~Szydłowski,
  %``Dynamics of the diffusive DM-DE interaction – Dynamical system approach,''
  JCAP {\bf 1607}, no. 07, 024 (2016)
  doi:10.1088/1475-7516/2016/07/024
  [arXiv:1603.07620 [gr-qc]].
  %%CITATION = doi:10.1088/1475-7516/2016/07/024;%%
  %14 citations counted in INSPIRE as of 11 May 2018
  \bibitem{Gao:2009me} 
  C.~Gao, M.~Kunz, A.~R.~Liddle and D.~Parkinson,
  %``Unified dark energy and dark matter from a scalar field different from quintessence,''
  Phys.\ Rev.\ D {\bf 81}, 043520 (2010)
  doi:10.1103/PhysRevD.81.043520
  [arXiv:0912.0949 [astro-ph.CO]].
  %%CITATION = doi:10.1103/PhysRevD.81.043520;%%
  %34 citations counted in INSPIRE as of 09 Dec 2018
  \bibitem{Benisty:2018qed} 
  D.~Benisty and E.~I.~Guendelman,
  %``Unified dark energy and dark matter from dynamical spacetime,''
  Phys.\ Rev.\ D {\bf 98}, no. 2, 023506 (2018)
  doi:10.1103/PhysRevD.98.023506
  [arXiv:1802.07981 [gr-qc]].
  %%CITATION = doi:10.1103/PhysRevD.98.023506;%%
  %1 citations counted in INSPIRE as of 10 Dec 2018
  \end{thebibliography}
\end{document}